# Robust multiuser detection in impulsive channels based on M-estimation using a new penalty function

M. Rastgou

In this paper, we consider the problem of multiuser detection in non-Gaussian channels. We propose a new penalty function for robust multiuser detection. The proposed detector outperforms other suboptimal detectors in non-Gaussian environment. Analytical and simulation result shows the performance of the proposed detector compare to other detectors.

*Introduction:* Detection and estimation theory is one of the core investigation areas in communications. Multiuser detection (MUD) is an estimation problem in which different users transmit their signals through the same channel in a multiple access channel (MAC). In this way the receiver explores desired signal through the ambient noise and interference generated by other user signals. Multiuser detection refers to detection techniques that aim at estimation of the desired signal in this circumstance. The maximum likelihood (ML) multiuser detection studied by Verdu [1], but the computational complexity of the ML detector exponentially increases by the number of users, hence using suboptimal detectors is of interest.

The majority of references has shown that in many cases the ambient noise model does not follow the Gaussian assumption [2, 3]. In this case the Gaussian assumption will cause a great performance loss in receiver [3]. On the other hand non-Gaussian noise distribution has a parametric model where estimation of these parameters is practically impossible. The problem of multiuser detection in non-Gaussian channels have been the subject of many references. Wang and Poor introduced a robust multiuser detection [2] based on Huber's robust Minimax detector [4]. They used some approximation to make the Minimax detector practical in non-Gaussian environment. Seyfe and Valaee [5] introduced a penalty function for robust multiuser detection, which outperforms the Minimax detector. In [6] and [7] Seyfe and Sharafat used Nonparametric statistics to introduce the Nonparametric multiuser detection.

In the next sections we introduce our new penalty function, we analyze the performance of this detector in different non-Gaussian environments and finally the simulation result shows the performance of this detector compares to other detectors.

*System model:* We assume a baseband synchronous direct sequence code division multiple access system (DS-CDMA) the received signal can be modeled in vector form as below

$$r = w + n = SAb + n \quad (1)$$

Where $r$ is a $N \times 1$ vector representing the sample vector of the received signal, $w$ refers to the signal vector of $K$ active users with dimension of $N$. $n$ represents the channel ambient noise sample vector. We assume that the noise samples are independent and identically (i.i.d) random variables. $S \triangleq [s_1, s_2, \dots, s_K]$ is matrix with each column containing the normalized signature vector of the corresponding user. $A$ is a diagonal $K \times K$ matrix with each diagonal entry specifying the corresponding user amplitude. $b$ depicts the column vector of BPSK symbol vector of users where $b_k = \pm 1, k = 1,2,\dots,K$ with equal probability. By substituting $\theta_k \triangleq A_k b_k$ then the received vector can be represented as

$$r = SAb + n = S\theta + n \quad (2)$$

With respect to this fact that $A_k$ is a positive scalar estimation of $b$ can be pursued by the estimate of $\theta$. The majority of references takes the well-recognized two term Gaussian mixture model as the ambient noise model which is an approximation to the Middleton class A noise model

$$f(x) = (1-\varepsilon)\mathcal{N}(0,\nu^2) + \varepsilon\mathcal{N}(0,\kappa\nu^2) \quad (3)$$

Where $\mathcal{N}(0,\nu^2)$ and $\mathcal{N}(0,\kappa\nu^2)$ terms represents the background and impulsive components of noise model. $\varepsilon$ represents the probability of which the impulsive components occur and $\kappa$ is the variance factor of the impulsive component. The two term Gaussian mixture model is a function of the parameters $\varepsilon$ and $\kappa$. Normally it is of interest to study the performance of detectors by varying the parameters $\varepsilon$ and $\kappa$ with fixed noise variance.

*The proposed detector:* It is well known that the least squares (LS) estimate heavily depends on the Gaussian noise assumption, a slight deviation from Gaussian noise model cause a substantial degradation of LS estimate. The LS estimate is a linear regression problem which aims at minimizing the Euclidean norm of the residual. Huber proposed to minimize the sum of a less rapidly increasing function of the residuals.

$$\hat{\boldsymbol{\theta}} = arg\, min_{\boldsymbol{\theta}} \sum_{j=1}^{N} \rho(r_j - w_j) \quad (4)$$

Consider $\rho$ has a derivative $\psi = \rho'$ then the equation (4) can be solved by

$$\sum_{j=1}^{N} \psi(r_j - w_j) s_j^k = 0, \quad k = 1,2,\dots,K \quad (5)$$

The function $\rho$ is called the penalty function and equation (4) has been named the M-estimator after maximum likelihood type estimator. Different penalty functions would lead to different detectors. For instance, taking $\rho = -\log f(x)$ results in $\psi = -\frac{f(x)}{f'(x)}$ where is the ML detector. Robustness of an estimator refers to the insensitivity of detector to small changes in the underlying statistical model. Huber proposed the Minimax detector, we propose a new derivative of penalty function for robust MUD as below

$$\psi_x = \begin{cases} \frac{x}{\sigma} & |x| \leq |\sigma| \\ \frac{\sigma}{x} & |x| > |\sigma| \end{cases} \quad (6)$$

Where $\sigma$ illustrates the standard deviation of noise distribution. We name equation (6) the *x*-detector after similarity to the Greek letter *x*. The MUD in non-Gaussian channels suffers heavily from impulses. This detector suppresses the impulsive components. The penalty function related to (6) would be

$$\rho_x = \begin{cases} \frac{x^2}{2\sigma} + C & |x| \leq |\sigma| \\ \sigma \ln x + C & |x| > |\sigma| \end{cases} \quad (7)$$

Where $C$ is a constant term, Equation (7) shows that the proposed penalty function is a nonlinear function which emphasizes on small values of the received samples for small amplitude it behaves like the LS estimator and it suppresses large component.

*Performance analysis:* Hampel showed that the robustness of the estimator is under the bounded and continuous $\psi$ function [8]. For the odd symmetric $\psi$ the estimator is consistent [1]. It is shown that the asymptotic variance of the estimation error of $\boldsymbol{\theta}$ at the noise distribution of $f(x)$ can be computed as [4]

$$V = \frac{\int \psi^2(x) f(x) dx}{[\int \psi'(x) f(x) dx]^2} \quad (8)$$

For the *x*-detector the asymptotic variance of the estimation error after some computation results in

$$\int \psi^2(x) f(x) dx = 2\left[\frac{\nu^2}{2\sigma^2}(1 + (\kappa-1)\varepsilon) - \left(\frac{\nu^2}{\sigma^2} - \frac{\sigma^2}{\nu^2}\right)(1-\varepsilon)Q\left(\frac{\sigma}{\nu}\right) - \left(\frac{\kappa\nu^2}{\sigma^2} - \frac{\sigma^2}{\kappa\nu^2}\right)\varepsilon Q\left(\frac{\sigma}{\sqrt{\kappa}\nu}\right) + \left(\frac{\sigma}{\nu} - \frac{\nu}{\sigma}\right)\frac{(1-\varepsilon)}{\sqrt{2\pi}}exp\left(-\frac{\sigma^2}{2\nu^2}\right) + \left(\frac{\sigma}{\nu} - \frac{\kappa\nu}{\sigma}\right)\frac{\varepsilon}{\sqrt{2\pi\kappa}}exp\left(-\frac{\sigma^2}{2\kappa\nu^2}\right)\right] \quad (9)$$

$$\int \psi'(x) f(x) dx = 2\left[\frac{1}{2\sigma} + \left(\frac{\sigma}{v^2} - \frac{1}{\sigma}\right)(1-\varepsilon)Q\left(\frac{\sigma}{v}\right) + \left(\frac{\sigma}{\kappa v^2} - \frac{1}{\sigma}\right)\varepsilon Q\left(\frac{\sigma}{\sqrt{\kappa}v}\right) - \frac{(1-\varepsilon)}{\sqrt{2\pi}v}\exp\left(-\frac{\sigma^2}{2v^2}\right) - \frac{\varepsilon}{\sqrt{2\pi\kappa}v}\exp\left(-\frac{\sigma^2}{2\kappa v^2}\right)\right] \quad (10)$$

We can compare the performance of the $x$-detector to other detectors by defining the asymptotic relative efficiency (ARE) as the ratio of the asymptotic variance of other detectors to that of the $x$-detector.

$$\text{ARE} = \frac{V}{V_x} \quad (11)$$

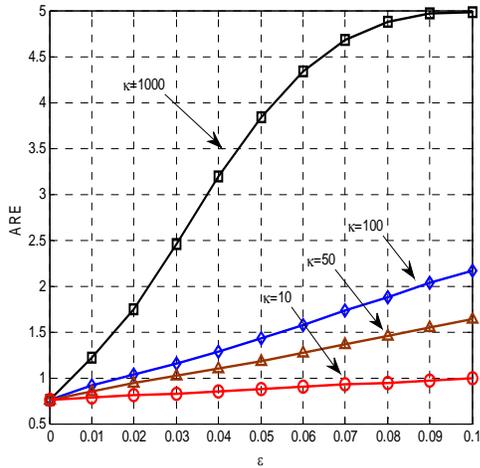

**Fig. 1** *The ARE of the x-detector to the Minimax detector as a function of $\varepsilon$ and $\kappa = 10, 50, 100, 1000$ with unit total noise variance i.e. $\sigma^2 \triangleq (1-\varepsilon)v^2 + \varepsilon\kappa v^2 = 1$.*

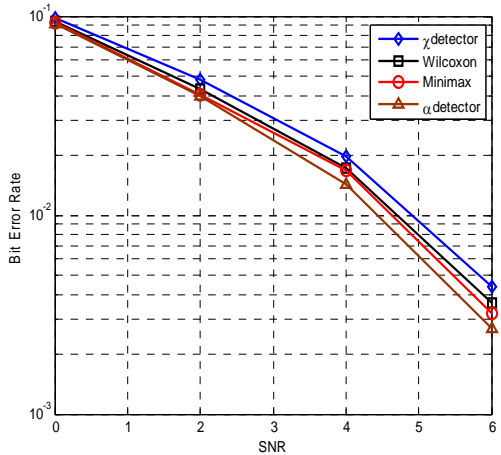

**Fig. 2** *BER against SNR of the user1 in a synchronous DS-CDMA with perfect power control for the Minimax detector, the α-detector, the nonparametric Wilcoxon detector and the x-detector where $K = 5$, $N = 31$, $\varepsilon = 0.01$ and $\kappa = 100$.*

Fig. 1 shows the performance of the $x$-detector compare to the Minimax detector it is evident that with increasing values of the parameters ε and κ the $x$-detector outperforms the Minimax detector.

*Simulation results:* We have considered a synchronous DS_CDMA system with $K = 5$ users where the spreading code of each user is a shifted m-sequence of length $N = 31$. We use the two term Gaussian mixture model as our non-Gaussian noise model. The signal to noise ratio is defined as the desired signal received power to the noise variance i.e. $\frac{A_k^2}{\sigma^2}$. We use an iterative Newton type algorithm to estimate the data vector. Assume $\hat{\boldsymbol{\theta}}^{(i)}$ as the estimate of $\boldsymbol{\theta}$ at the $i_{th}$ step then $\hat{\boldsymbol{\theta}}^{(i+1)}$ will be

$$\hat{\boldsymbol{\theta}}^{(i+1)} = \hat{\boldsymbol{\theta}}^{(i)} - \mu (\boldsymbol{S}^T\boldsymbol{S})^{-1}\boldsymbol{S}^T \psi_x(\boldsymbol{r} - \boldsymbol{S}\hat{\boldsymbol{\theta}}^{(i)}) \quad (12)$$

Where $\mu$ is the step-size of the algorithm, for the bounded $\psi'$ the algorithm will be convergent [4].

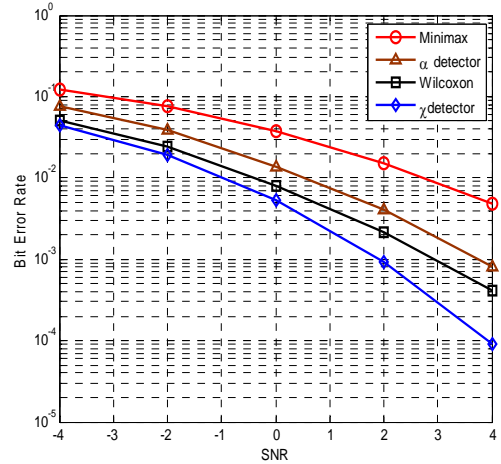

**Fig. 3** *BER against SNR of the user1 in a synchronous DS-CDMA with perfect power control for the Minimax detector, the α-detector, the nonparametric Wilcoxon detector and the x-detector where $K = 5$, $N = 31$, $\varepsilon = 0.1$ and $\kappa = 100$.*

Fig. 2 and Fig. 3 illustrates the BER of the $x$-detector compares to other detectors it can be seen that at $\varepsilon = 0.1$ the $x$-detector shows comparable performance compare to other detectors at $\varepsilon = 0.1$ it shows better performance in which it has 4 and 2dB gain over the Minimax detector and the α-detector respectively it has to be mentioned that nonparametric detector suffers from greater complexity.

*Conclusion:* we introduced a new robust multiuser detection base on M-estimation in impulsive channels since the ML detector suffers from more computational complexity, it is of interest to use suboptimal detectors we showed that our proposed detector outperforms other suboptimal detectors in impulsive channels.

M. Rastgou (*Engineering Department, Shahed University, Tehran, Iran*)
E-mail: